\documentclass[11pt,a4paper]{article}
\usepackage[margin=0.82in]{geometry}
\usepackage{amsmath,amssymb,bm}
\usepackage{graphicx}
\usepackage{booktabs,tabularx}
\usepackage{siunitx}
\usepackage{hyperref}
\usepackage{microtype}
\usepackage{caption}
\usepackage{float}
\usepackage{enumitem}
\hypersetup{colorlinks=true,linkcolor=blue,citecolor=blue,urlcolor=blue}
\graphicspath{{figures/}}
\title{\textbf{Probing CPT Violation with High-Energy Neutrinos: Channel-Resolved Oscillations in Earth Matter from TeV to PeV}}
\author{Anurag Sharma\and Bipin Singh Koranga\thanks{Corresponding author: bskoranga@kmc.du.ac.in}\and Vivek Kumar Nautiyal\\[3pt]
\small $^{1}$Department of Physics, Kirori Mal College, University of Delhi, Delhi 110007, India\\
\small $^{2}$Department of Physics, Chaudhary Charan Singh University, Meerut 250004, India}
\date{}
\begin{document}
\maketitle
\begin{abstract}
We present a numerical three-flavour study of neutrino and antineutrino propagation through a four-shell Earth model in the presence of a CPT-odd contribution to the effective Hamiltonian. The calculation combines the standard PMNS mixing framework, vacuum mass-squared terms, the charged-current matter potential, and a Hermitian CPT-odd matrix. Ordered evolution through concentric Earth layers is used to evaluate all nine flavour-transition channels, the normalized same-index neutrino--antineutrino channel asymmetry, and the zenith-dependent detector flavour composition for pion-decay, muon-damped, and neutron-decay source scenarios. Internal checks of PMNS unitarity, evolution-operator unitarity, and probability conservation are satisfied at approximately machine precision. For vertical incidence at 100 TeV, the probability matrix agrees with the benchmark values to the quoted precision. Over the tested high-energy interval, the logarithmic slope of the absolute normalized $e\to\mu$ channel asymmetry is approximately $-1$, consistent with the leading $1/E$ scaling expected when the vacuum term is a perturbation to energy-independent matter and CPT-odd terms. The detector-level flavour scan reproduces the broad zenith dependence, while some muon-flavour extrema remain discrepant from the rounded benchmark values. We retain and discuss this discrepancy rather than tune the benchmark parameter to force agreement.
\end{abstract}

\section{Introduction}
CPT symmetry is a cornerstone of local, Lorentz-invariant quantum field theory. Under the assumptions of the CPT theorem, violation of CPT entails violation of Lorentz invariance [6,35]. Neutrino oscillations provide a sensitive setting in which small differences in propagation eigenvalues can accumulate over long baselines; see Ref.~[37] for a review of massive-neutrino phenomenology. Phenomenological and Standard-Model Extension (SME) descriptions of Lorentz- and CPT-violating neutrino propagation have been developed in Refs.~[4,5,7--13,18,19,21--23,36]. Experimental searches using atmospheric neutrinos, including Super-Kamiokande and IceCube samples, constrain SME coefficients in specific operator frameworks [24,25]. The sensitivity of high-energy atmospheric samples to related effects has also been studied [26,27]. These limits should not be identified directly with the effective splitting parameter used in the restricted ansatz studied here.

Matter effects complicate the interpretation of neutrino--antineutrino probability differences. Even in a CPT-invariant fundamental theory, propagation through ordinary matter can generate an extrinsic neutrino--antineutrino asymmetry [17]. The distinction between intrinsic CPT-odd contributions and matter-induced effects is therefore important when interpreting channel-by-channel comparisons. Earlier phenomenological studies considered CPT-odd effects in atmospheric and long-baseline oscillations [8--10,15--17,20]. Astrophysical neutrinos provide a complementary setting: flavour mixing over cosmic baselines maps source compositions into surface compositions [28--30,38,39], while subsequent Earth propagation modifies the flavour fractions for up-going trajectories.

The present work is a numerical implementation and reproducibility study of a three-flavour propagation framework, following the parameter choices of an earlier benchmark calculation supplied by the authors [42]. Our focus is (i) all nine flavour-transition channels, (ii) the energy dependence of a normalized same-index neutrino--antineutrino channel asymmetry, and (iii) source-dependent flavour fractions after propagation through the Earth. The benchmark framework and our independent numerical implementation are distinguished explicitly. In particular, the detector-level comparison retains residual differences in some muon-flavour extrema.

\section{Theoretical framework}
\subsection{Effective Hamiltonians}
In the flavour basis, the neutrino Hamiltonian is
\begin{equation}
H_\nu=U\,\mathrm{diag}\!\left(0,\frac{\Delta m^2_{21}}{2E},\frac{\Delta m^2_{31}}{2E}\right)U^\dagger+b+\mathrm{diag}(V,0,0),
\label{eq:Hnu}
\end{equation}
and the antineutrino Hamiltonian is
\begin{equation}
H_{\bar\nu}=U^*\,\mathrm{diag}\!\left(0,\frac{\Delta m^2_{21}}{2E},\frac{\Delta m^2_{31}}{2E}\right)U^T-b^*-\mathrm{diag}(V,0,0).
\label{eq:Hnubar}
\end{equation}
For the benchmark calculation, the Hermitian CPT-odd term is restricted to
\begin{equation}
b=U\,\mathrm{diag}(0,\delta b_{21},\delta b_{31})\,U^\dagger.
\label{eq:b}
\end{equation}
This ansatz uses the same mixing matrix for the mass and CPT-odd terms. It is a restricted phenomenological choice, not the most general Hermitian matrix.

\subsection{Mixing matrix and matter potential}
We use the standard PMNS parametrization [31,32], with $s_{ij}=\sin\theta_{ij}$, $c_{ij}=\cos\theta_{ij}$, and Dirac phase $\delta$:
\begin{equation}
U=\begin{pmatrix}
c_{12}c_{13}&s_{12}c_{13}&s_{13}e^{-i\delta}\\
-s_{12}c_{23}-c_{12}s_{23}s_{13}e^{i\delta}&c_{12}c_{23}-s_{12}s_{23}s_{13}e^{i\delta}&s_{23}c_{13}\\
s_{12}s_{23}-c_{12}c_{23}s_{13}e^{i\delta}&-c_{12}s_{23}-s_{12}c_{23}s_{13}e^{i\delta}&c_{23}c_{13}
\end{pmatrix}.
\end{equation}
The numerical inputs are $\sin^2\theta_{12}=0.307$, $\sin^2\theta_{13}=0.022$, $\sin^2\theta_{23}=0.55$, $\delta_{\rm CP}=195^\circ$, $\Delta m^2_{21}=7.4\times10^{-5}\,\mathrm{eV}^2$, and $\Delta m^2_{31}=2.5\times10^{-3}\,\mathrm{eV}^2$ (normal ordering), consistent with representative global-fit values [33,41].

The charged-current matter potential is [1,2]
\begin{equation}
V=\sqrt{2}G_F N_e=7.63\times10^{-14}Y_e\rho\ \mathrm{eV},
\end{equation}
where $\rho$ is in $\mathrm{g\,cm^{-3}}$. The electron fractions used below are model inputs adopted for the crust/mantle and core layers; they are stated as assumptions of this calculation rather than as values fixed by the PREM density model [3].

\subsection{Layer evolution and channel asymmetry}
For a constant-density layer $j$ of length $L_j$,
\begin{equation}
S_j=\exp\!\left[-iH_j\frac{L_j}{\hbar c}\right],\qquad S_{\rm tot}=S_nS_{n-1}\cdots S_1,
\end{equation}
where the product is ordered along the direction of propagation. We define
\begin{equation}
P_\nu^{\alpha\beta}=P(\nu_\alpha\to\nu_\beta)=\left|(S_{\rm tot}^{\nu})_{\beta\alpha}\right|^2,
\qquad
P_{\bar\nu}^{\alpha\beta}=P(\bar\nu_\alpha\to\bar\nu_\beta)=\left|(S_{\rm tot}^{\bar\nu})_{\beta\alpha}\right|^2.
\label{eq:prob}
\end{equation}
The normalized same-index channel asymmetry studied in this work is
\begin{equation}
A_{\nu\bar\nu}^{\alpha\beta}=\frac{P_\nu^{\alpha\beta}-P_{\bar\nu}^{\alpha\beta}}{P_\nu^{\alpha\beta}+P_{\bar\nu}^{\alpha\beta}}.
\label{eq:asym}
\end{equation}
It is distinct from the unnormalized probability difference. For $\alpha\ne\beta$, the same-index comparison in Eq.~(\ref{eq:asym}) is not the strictly CPT-conjugate comparison, which pairs $P(\nu_\alpha\to\nu_\beta)$ with $P(\bar\nu_\beta\to\bar\nu_\alpha)$ [14,17]. We use the same-index convention consistently for the numerical curves and slope quoted below. Consequently, $A_{\nu\bar\nu}^{\alpha\beta}$ in matter is a channel asymmetry and should not be interpreted by itself as a clean, matter-independent measurement of fundamental CPT violation.

When the CPT-even vacuum term is neglected, Eqs.~(\ref{eq:Hnu}) and (\ref{eq:Hnubar}) give $H_{\bar\nu}=-H_\nu^*$, so the corresponding same-index probabilities are equal. At finite vacuum contribution, interference with the energy-independent matter and CPT-odd terms can generate a nonzero asymmetry. In the high-energy regime where the vacuum term is perturbative, a leading behaviour proportional to $1/E$ is expected within this model.

\section{Numerical setup and method}
The Earth is represented by four concentric constant-density shells. The adopted densities are $2.8$, $4.5$, $11.0$, and $13.0\,\mathrm{g\,cm^{-3}}$ for crust, mantle, outer core, and inner core, respectively, following a PREM-inspired approximation [3]. We use $Y_e=0.4957$ in crust and mantle and $Y_e=0.4656$ in the core. The radial boundaries are $R_\oplus=6371$ km, crust bottom $6336$ km, core--mantle boundary $3480$ km, and inner-core boundary $1221$ km. The benchmark describes a four-shell PREM-inspired model but does not tabulate every boundary in the source text available to us. These radii are therefore the explicit geometry adopted in our implementation and are not claimed to be the benchmark authors' exact internal choices.

Two benchmark cases are considered: (A) $\delta b_{31}\ne0$, $\delta b_{21}=0$; and (B) $\delta b_{21}\ne0$, $\delta b_{31}=0$. The illustrative splitting used for these channel studies is $5\times10^{-23}$ GeV. This scale is comparable in order of magnitude to parameters discussed in earlier atmospheric-neutrino studies [15,16], but no direct conversion to a current experimental limit is claimed. The detector-composition scan uses $E=100$ TeV and $\delta b_{31}=2\times10^{-22}$ GeV.

For each zenith angle, the impact parameter and chord lengths through the shells are calculated. The path consists of far-side and near-side segments in the order crust, mantle, outer core, inner core, inner core, outer core, mantle, crust, with core segments present only for sufficiently vertical trajectories. The layer exponentials are evaluated using \texttt{scipy.linalg.expm} and multiplied in propagation order. We use $\hbar c=1.973269804\times10^{-10}\,\mathrm{eV\,km}$. The probability matrices are stored with initial flavour in rows and final flavour in columns.

\section{Results}
\subsection{Numerical validation}
For vertical incidence at $E=100$ TeV in scenario A, the calculated probability matrix is
\begin{equation}
P=\begin{pmatrix}
0.99231588&0.00422631&0.00345781\\
0.00422623&0.00958235&0.98619142\\
0.00345789&0.98619134&0.01035077
\end{pmatrix}.
\label{eq:P100}
\end{equation}
The PMNS unitarity error is $1.11\times10^{-16}$; the neutrino and antineutrino evolution-operator unitarity errors are approximately $1.11\times10^{-15}$; and the probability-conservation errors are approximately $10^{-15}$. The vertical chord length is 12742 km. These checks establish internal numerical consistency. The matrix in Eq.~(\ref{eq:P100}) agrees with the rounded benchmark values to the quoted precision [42].

\begin{figure}[H]\centering
\includegraphics[width=0.96\linewidth]{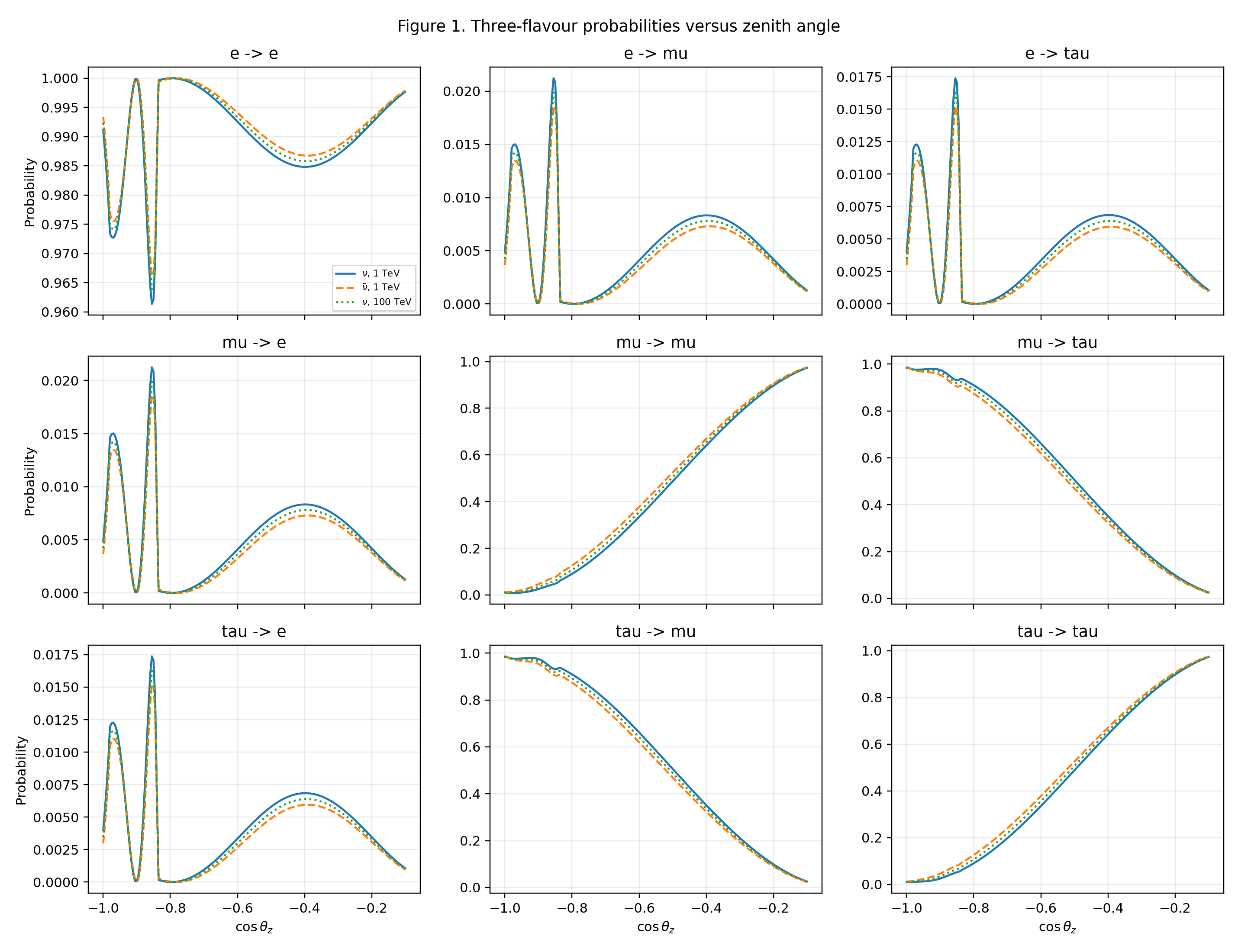}
\caption{Three-flavour transition probabilities $P(\nu_\alpha\to\nu_\beta)$ versus $\cos\theta_z$ for scenario A, with $\delta b_{31}=5\times10^{-23}$ GeV. Each panel corresponds to one channel; curves show neutrinos and antineutrinos at 1 TeV and neutrinos at 100 TeV.}\label{fig:prob}
\end{figure}

Figure~\ref{fig:prob} shows all nine transition channels versus zenith angle for neutrinos and antineutrinos at 1 TeV and neutrinos at 100 TeV. The sharp structures near $\cos\theta_z\simeq-0.84$ occur close to the core--mantle-boundary tangency, at
\begin{equation}
\cos\theta_z=-\sqrt{1-(R_{\rm CMB}/R_\oplus)^2}\simeq-0.838.
\end{equation}

\subsection{Energy dependence}
The absolute normalized same-index asymmetry $|A_{\nu\bar\nu}^{\alpha\beta}|$ was evaluated at vertical incidence from $10^2$ to $10^6$ GeV. A log--log fit to the $e\to\mu$ channel over 190 TeV--1 PeV gives
\begin{equation}
\frac{d\log_{10}|A_{\nu\bar\nu}^{e\mu}|}{d\log_{10}E}=-0.99999992.
\end{equation}
The fitted slope is consistent with $-1$ to within $10^{-7}$ for the selected scan and fit window. The small residual from exactly $-1$ depends on numerical sampling and the fit interval and is not assigned separate physical significance. The leading scaling follows from the vacuum term decreasing as $1/E$ relative to the energy-independent terms in the Hamiltonian.
\begin{figure}[H]\centering
\includegraphics[width=0.78\linewidth]{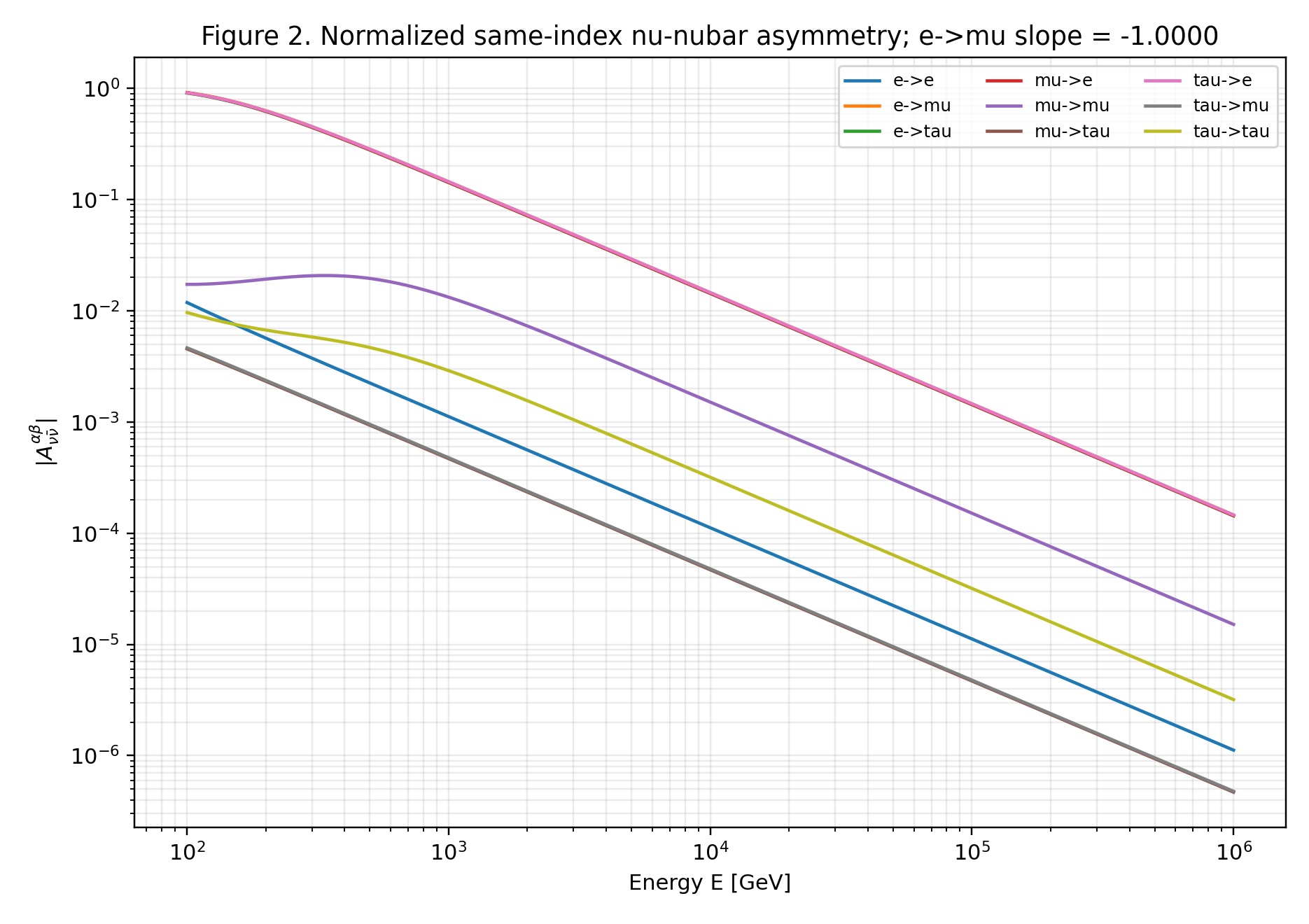}
\caption{Absolute normalized same-index neutrino--antineutrino channel asymmetry for all nine channels versus energy at vertical incidence in scenario A. Both axes are logarithmic.}\label{fig:energy}
\end{figure}

\subsection{Dependence on the CPT-odd splitting}
\begin{figure}[H]\centering
\includegraphics[width=0.96\linewidth]{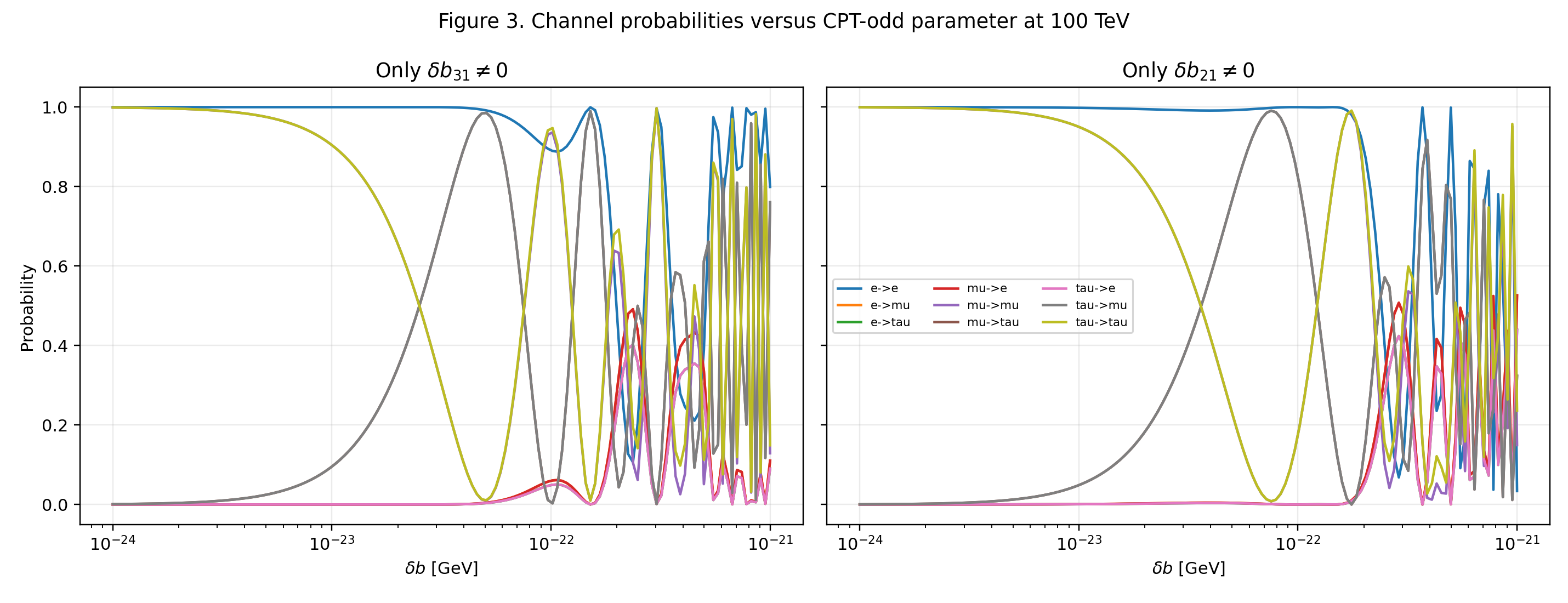}
\caption{Channel probabilities versus CPT-odd splitting at 100 TeV and vertical incidence for scenario A (left, only $\delta b_{31}\ne0$) and scenario B (right, only $\delta b_{21}\ne0$).}\label{fig:split}
\end{figure}

Figure~\ref{fig:split} shows the channel probabilities at 100 TeV as a function of the splitting for scenarios A and B. For a vertical chord of 12742 km, the phase estimate is $\delta b L/(\hbar c)\simeq0.65$ for $\delta b=10^{-23}$ GeV and approximately $6.5$ for $10^{-22}$ GeV. Thus, a splitting of order $10^{-23}$ GeV can already produce a non-negligible phase, while oscillatory dependence becomes pronounced around $10^{-22}$ GeV. This phase estimate is more informative than describing the propagation as indistinguishable from the zero-splitting case throughout that range.

\section{Detector-level flavour composition}
For an astrophysical source composition $f_\alpha^{\rm src}$, the vacuum-averaged flavour fraction at the Earth surface is
\begin{equation}
f_\beta=\sum_{\alpha,i}f_\alpha^{\rm src}|U_{\alpha i}|^2|U_{\beta i}|^2.
\end{equation}
After propagation through the Earth, the detector-level fraction is
\begin{equation}
R_\beta(\cos\theta_z)=\sum_\alpha f_\alpha P_{\alpha\beta}(\cos\theta_z),
\end{equation}
where $f_\alpha$ denotes the surface composition and $P_{\alpha\beta}$ is the neutrino probability matrix in the row-initial, column-final convention. The calculated surface fractions $(f_e,f_\mu,f_\tau)$ are $(0.304043,0.354534,0.341422)$ for pion decay $(1:2:0)$, $(0.181074,0.441265,0.377661)$ for a muon-damped source $(0:1:0)$, and $(0.549982,0.181074,0.268944)$ for neutron decay $(1:0:0)$.

\begin{figure}[H]\centering
\includegraphics[width=0.99\linewidth]{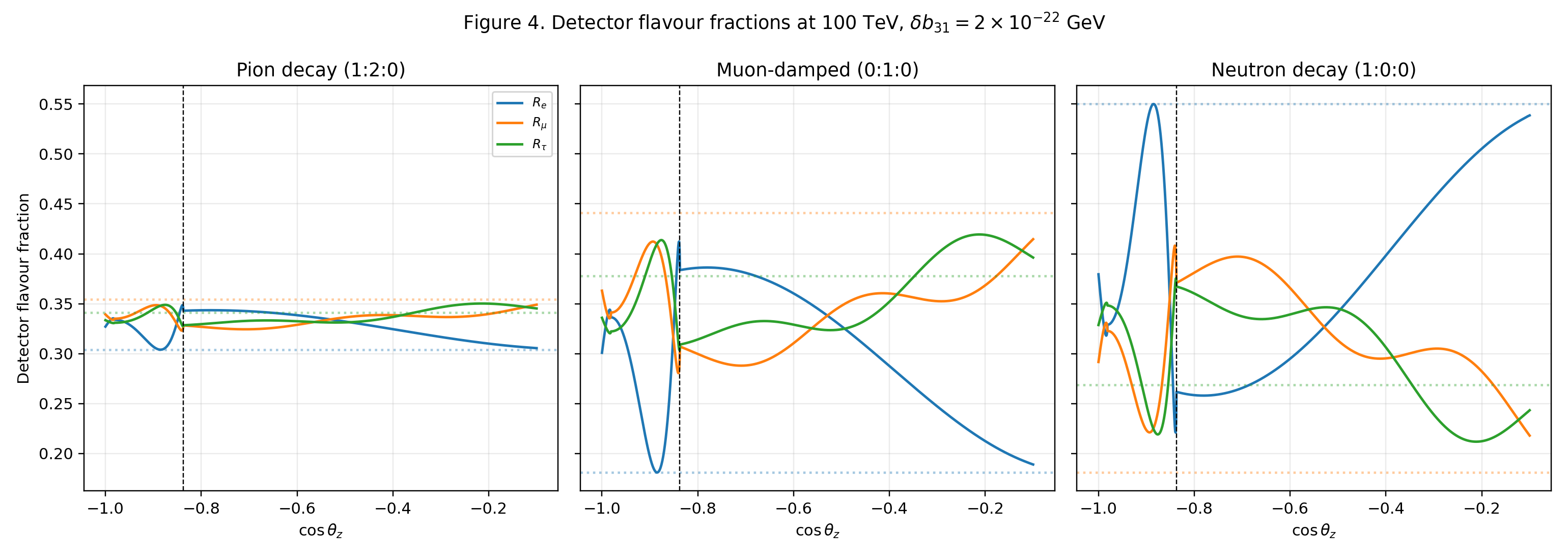}
\caption{Detector-level flavour fractions versus zenith angle at 100 TeV for pion-decay $(1:2:0)$, muon-damped $(0:1:0)$, and neutron-decay $(1:0:0)$ source compositions, with $\delta b_{31}=2\times10^{-22}$ GeV. The vertical dashed line marks the core--mantle-boundary tangency; dotted lines indicate the corresponding surface fractions.}\label{fig:detector}
\end{figure}

Figure~\ref{fig:detector} shows $R_\beta$ over $-1.0\leq\cos\theta_z\leq-0.1$, sampled at 1001 equally spaced points, for $E=100$ TeV and $\delta b_{31}=2\times10^{-22}$ GeV. Table~\ref{tab:extrema} compares the extrema with the rounded ranges reported in the benchmark [42]. The electron-flavour ranges are close. Several muon-flavour extrema differ by more than the rounding precision, most notably the upper endpoint for the muon-damped source and the lower endpoint for the neutron-decay source. We do not tune the stated CPT-odd coefficient to remove these differences.
\begin{table}[H]\centering\small
\caption{Minimum--maximum ranges of detector flavour fractions over $-1.0\leq\cos\theta_z\leq-0.1$, comparing this calculation with the rounded benchmark ranges at $\delta b_{31}=2\times10^{-22}$ GeV.}\label{tab:extrema}
\begin{tabular}{lcccc}\toprule
Source & $R_e$ (this work) & $R_e$ (benchmark) & $R_\mu$ (this work) & $R_\mu$ (benchmark)\\\midrule
Pion decay & 0.304--0.349 & 0.304--0.348 & 0.323--0.349 & 0.323--0.353\\
Muon-damped & 0.181--0.412 & 0.181--0.411 & 0.281--0.415 & 0.283--0.434\\
Neutron decay & 0.221--0.550 & 0.222--0.550 & 0.218--0.408 & 0.191--0.405\\\bottomrule
\end{tabular}
\end{table}

\section{Discussion}
The implementation reproduces the principal numerical structure of the benchmark framework: PMNS unitarity, unitary layer evolution, probability conservation, a high-energy same-index channel-asymmetry slope close to $-1$, and a vertical 100 TeV probability matrix matching the benchmark values to the displayed precision. The detector-level scan captures the broad source-dependent zenith variation. However, several extrema, particularly in $R_\mu$, remain discrepant. The benchmark text available for this comparison does not specify every shell boundary or the complete angular sampling and source-to-surface averaging prescription. The adopted geometry and 1001-point scan are therefore stated explicitly, and pointwise agreement is not claimed.

The same-index quantity in Eq.~(\ref{eq:asym}) must also be distinguished from the strictly CPT-conjugate transition comparison for off-diagonal channels. Matter can generate extrinsic neutrino--antineutrino differences [17], so the normalized channel asymmetry is not by itself an unambiguous fundamental CPT observable. The restricted matrix ansatz in Eq.~(\ref{eq:b}) further limits the scope of the conclusions.

The calculation does not include neutrino attenuation in the Earth, tau regeneration, detector response, event reconstruction, or statistical uncertainties. These effects become increasingly relevant toward the upper end of the TeV--PeV interval [34,40]. The plotted quantities are propagated flavour fractions, not event rates or experimental sensitivity forecasts. The benchmark values of $\delta b$ have not been converted into or compared quantitatively with current experimental SME limits [14,24,25].

\section{Conclusions}
We have implemented three-flavour neutrino and antineutrino propagation through a four-shell Earth model with vacuum, matter, and CPT-odd Hamiltonian terms. The calculation evaluates all nine transition channels, a normalized same-index neutrino--antineutrino channel asymmetry, and source-dependent detector flavour fractions. Internal unitarity and probability-conservation checks are satisfied at approximately machine precision. The vertical 100 TeV probability matrix agrees with the benchmark to the quoted precision, and the fitted high-energy $e\to\mu$ channel-asymmetry slope is approximately $-1$. The detector-level calculation reproduces the broad zenith dependence, but some muon-flavour extrema remain discrepant. These differences are reported transparently and are not removed by tuning the stated CPT-odd coefficient. A more complete Earth model, explicit benchmark geometry and sampling details, and detector-level effects are required for a more definitive comparison.

\section*{Data and code availability}
The Python implementation and machine-readable numerical diagnostics are included as ancillary files in the source package accompanying this manuscript. The package does not claim a permanent public repository DOI; a repository identifier should be added only after a public versioned release has been created.


\begin{thebibliography}{99}
\bibitem{wolf} L. Wolfenstein, ``Neutrino oscillations in matter,'' Phys. Rev. D \textbf{17} (1978) 2369--2374. doi:10.1103/PhysRevD.17.2369.
\bibitem{msw} S. P. Mikheyev and A. Yu. Smirnov, ``Resonance enhancement of oscillations in matter and solar neutrino spectroscopy,'' Sov. J. Nucl. Phys. \textbf{42} (1985) 913--917.
\bibitem{prem} A. M. Dziewonski and D. L. Anderson, ``Preliminary reference Earth model,'' Phys. Earth Planet. Inter. \textbf{25} (1981) 297--356. doi:10.1016/0031-9201(81)90046-7.
\bibitem{ck97} D. Colladay and V. A. Kosteleck\'y, ``CPT violation and the standard model,'' Phys. Rev. D \textbf{55} (1997) 6760--6774. doi:10.1103/PhysRevD.55.6760; arXiv:hep-ph/9703464.
\bibitem{ck98} D. Colladay and V. A. Kosteleck\'y, ``Lorentz-violating extension of the standard model,'' Phys. Rev. D \textbf{58} (1998) 116002. doi:10.1103/PhysRevD.58.116002; arXiv:hep-ph/9809521.
\bibitem{greenberg} O. W. Greenberg, ``CPT violation implies violation of Lorentz invariance,'' Phys. Rev. Lett. \textbf{89} (2002) 231602. arXiv:hep-ph/0201258.
\bibitem{kostsam} V. A. Kosteleck\'y and S. Samuel, ``Spontaneous breaking of Lorentz symmetry in string theory,'' Phys. Rev. D \textbf{39} (1989) 683--685.
\bibitem{murayama} H. Murayama and T. Yanagida, ``LSND, SN1987A, and CPT violation,'' Phys. Lett. B \textbf{520} (2001) 263--268. arXiv:hep-ph/0010178.
\bibitem{barenboim02} G. Barenboim, L. Borissov, J. Lykken and A. Yu. Smirnov, ``Neutrinos as the messengers of CPT violation,'' JHEP \textbf{10} (2002) 001. arXiv:hep-ph/0108199.
\bibitem{barenboim03} G. Barenboim and J. Lykken, ``A model of CPT violation for neutrinos,'' Phys. Lett. B \textbf{554} (2003) 73--80. arXiv:hep-ph/0210411.
\bibitem{km04} V. A. Kosteleck\'y and M. Mewes, ``Lorentz and CPT violation in neutrinos,'' Phys. Rev. D \textbf{69} (2004) 016005. doi:10.1103/PhysRevD.69.016005; arXiv:hep-ph/0309025.
\bibitem{diaz09} J. S. D\'iaz, V. A. Kosteleck\'y and M. Mewes, ``Perturbative Lorentz and CPT violation for neutrino and antineutrino oscillations,'' Phys. Rev. D \textbf{80} (2009) 076007. arXiv:0908.1401.
\bibitem{km12} V. A. Kosteleck\'y and M. Mewes, ``Neutrinos with Lorentz-violating operators of arbitrary dimension,'' Phys. Rev. D \textbf{85} (2012) 096005. arXiv:1112.6395.
\bibitem{kr11} V. A. Kosteleck\'y and N. Russell, ``Data tables for Lorentz and CPT violation,'' Rev. Mod. Phys. \textbf{83} (2011) 11--31. doi:10.1103/RevModPhys.83.11; arXiv:0801.0287.
\bibitem{datta04} A. Datta, R. Gandhi, P. Mehta and S. Uma Sankar, ``Atmospheric neutrinos as a probe of CPT violation,'' Phys. Lett. B \textbf{597} (2004) 356--361. doi:10.1016/j.physletb.2004.07.035; arXiv:hep-ph/0312027.
\bibitem{samanta10} A. Samanta, ``Probing CPT violation in neutrino oscillation: A three flavor analysis,'' Phys. Lett. B \textbf{693} (2010) 296--300. doi:10.1016/j.physletb.2010.08.062; arXiv:1005.4851.
\bibitem{jacobson04} M. Jacobson and T. Ohlsson, ``Extrinsic CPT violation in neutrino oscillations in matter,'' Phys. Rev. D \textbf{69} (2004) 013003. doi:10.1103/PhysRevD.69.013003; arXiv:hep-ph/0305064.
\bibitem{koranga06} B. S. Koranga, M. Narayan and S. Uma Sankar, ``Possible CPT violation from Planck scale effects,'' arXiv:hep-ph/0607274.
\bibitem{koranga11} B. S. Koranga, M. Narayan and S. Uma Sankar, ``Relation between CPT violation in neutrino masses and mixings,'' Int. J. Theor. Phys. \textbf{50} (2011) 760--766. doi:10.1007/s10773-010-0612-y; arXiv:0912.5005.
\bibitem{koranga14} B. S. Koranga and P. Khurana, ``CPT violation in atmospheric neutrino oscillation: A two flavour matter effects,'' Int. J. Theor. Phys. \textbf{53} (2014) 3737--3743. doi:10.1007/s10773-014-2126-5.
\bibitem{pandit26} T. Pandit, B. S. Koranga, A. Pant and V. K. Nautiyal, ``Matter-induced CPT violation and Earth-density stratification effects in long-baseline neutrino oscillation experiments,'' arXiv:2608.00436.
\bibitem{koranga26} B. S. Koranga, benchmark manuscript on three-flavour TeV--PeV neutrino oscillations with Earth matter effects, dated 17 November 2026, supplied to the authors for comparison (unpublished manuscript).
\bibitem{korangakhan26} B. S. Koranga and I. Khan, ``Planck-scale signatures in vacuum neutrino oscillations,'' arXiv:2609.06203.
\bibitem{farooq26} B. W. Farooq, B. S. Koranga and A. Chatterjee, ``Quantum Fisher information as a probe of sterile neutrino new physics: geometric advantage of KM3NeT over IceCube,'' arXiv:2604.01256.
\bibitem{superk15} K. Abe et al. (Super-Kamiokande Collaboration), ``Test of Lorentz invariance with atmospheric neutrinos,'' Phys. Rev. D \textbf{91} (2015) 052003. arXiv:1410.4267.
\bibitem{icecube18} M. G. Aartsen et al. (IceCube Collaboration), ``Neutrino interferometry for high-precision tests of Lorentz symmetry with IceCube,'' Nature Phys. \textbf{14} (2018) 961--966. doi:10.1038/s41567-018-0172-2; arXiv:1709.03434.
\bibitem{hooper05} D. Hooper, D. Morgan and E. Winstanley, ``Lorentz and CPT invariance violation in high-energy neutrinos,'' Phys. Rev. D \textbf{72} (2005) 065009.
\bibitem{gonzalez05} M. C. Gonzalez-Garcia, F. Halzen and M. Maltoni, ``Physics reach of high-energy and high-statistics IceCube atmospheric neutrino data,'' Phys. Rev. D \textbf{71} (2005) 093010. arXiv:hep-ph/0502223.
\bibitem{learned95} J. G. Learned and S. Pakvasa, ``Detecting $\nu_\tau$ oscillations at PeV energies,'' Astropart. Phys. \textbf{3} (1995) 267--274. doi:10.1016/0927-6505(94)00043-3; arXiv:hep-ph/9405296.
\bibitem{athar00} H. Athar, M. Jezabek and O. Yasuda, ``Effects of neutrino mixing on high-energy cosmic neutrino flux,'' Phys. Rev. D \textbf{62} (2000) 103007. doi:10.1103/PhysRevD.62.103007; arXiv:hep-ph/0005104.
\bibitem{barenboim03cosmic} G. Barenboim and C. Quigg, ``Neutrino observatories can characterize cosmic sources and neutrino properties,'' Phys. Rev. D \textbf{67} (2003) 073024. arXiv:hep-ph/0301220.
\bibitem{mns62} Z. Maki, M. Nakagawa and S. Sakata, ``Remarks on the unified model of elementary particles,'' Prog. Theor. Phys. \textbf{28} (1962) 870--880. doi:10.1143/PTP.28.870.
\bibitem{pdg22} R. L. Workman et al. (Particle Data Group), ``Review of Particle Physics,'' Prog. Theor. Exp. Phys. (2022) 083C01. doi:10.1093/ptep/ptac097.
\bibitem{esteban20} I. Esteban, M. C. Gonzalez-Garcia, M. Maltoni, T. Schwetz and A. Zhou, ``The fate of hints: updated global analysis of three-flavour neutrino oscillations,'' JHEP \textbf{09} (2020) 178. doi:10.1007/JHEP09(2020)178; arXiv:2007.14792.
\bibitem{gandhi96} R. Gandhi, C. Quigg, M. H. Reno and I. Sarcevic, ``Ultrahigh-energy neutrino interactions,'' Astropart. Phys. \textbf{5} (1996) 81--110. doi:10.1016/0927-6505(96)00008-4; arXiv:hep-ph/9512364.
\bibitem{luders57} G. L\"uders, ``Proof of the TCP theorem,'' Ann. Phys. \textbf{2} (1957) 1--15.
\bibitem{diaz12} J. S. D\'iaz and V. A. Kosteleck\'y, ``Lorentz- and CPT-violating models for neutrino oscillations,'' Phys. Rev. D \textbf{85} (2012) 016013. doi:10.1103/PhysRevD.85.016013; arXiv:1108.1799.
\bibitem{gonzalez08} M. C. Gonzalez-Garcia and M. Maltoni, ``Phenomenology with massive neutrinos,'' Phys. Rep. \textbf{460} (2008) 1--129. arXiv:0704.1800.
\bibitem{bustamante15} M. Bustamante, J. F. Beacom and W. Winter, ``Theoretically palatable flavor combinations of astrophysical neutrinos,'' Phys. Rev. Lett. \textbf{115} (2015) 161302. arXiv:1506.02645.
\bibitem{arguelles15} C. A. Arg\"uelles, T. Katori and J. Salvado, ``New physics in astrophysical neutrino flavor,'' Phys. Rev. Lett. \textbf{115} (2015) 161303. arXiv:1506.02043.
\bibitem{vincent17} A. C. Vincent, C. A. Arg\"uelles and A. Kheirandish, ``High-energy neutrino attenuation in the Earth and its associated uncertainties,'' JCAP \textbf{11} (2017) 012. arXiv:1706.09895.
\bibitem{nufit24} I. Esteban, M. C. Gonzalez-Garcia, M. Maltoni, I. Martinez-Soler, J. P. Pinheiro and T. Schwetz, ``NuFit-6.0: updated global analysis of three-flavor neutrino oscillations,'' JHEP \textbf{12} (2024) 216. arXiv:2410.05380.
\end{thebibliography}
\end{document}